\documentclass[prb,twocolumn,aps,showpacs,preprintnumbers]{revtex4}
\usepackage{graphicx}
\usepackage{dcolumn}
\usepackage{bm}

\begin{document}


\title{Statistics of Transmission Eigenvalues for a Disordered Quantum
Point Contact}

\author{G.~Campagnano, O.~N.~Jouravlev, Ya.~M.~Blanter, and
Yu.~V.~Nazarov} 
 \affiliation{Department of NanoScience and DIMES,
Delft University of Technology, Lorentzweg 1, 2628 CJ Delft, The
Netherlands  }
\date{\today}

\begin{abstract}
We study the distribution of transmission eigenvalues of a quantum
point contact with nearby impurities. In the semi-classical case (the
chemical potential lies at the conductance plateau) we find that the
transmission properties of this system are obtained from the ensemble
of Gaussian random reflection matrices. The distribution only depends
on the number of open transport channels and the average reflection
eigenvalue and crosses over from the Poissonian for one open channel to the
form predicted by the circuit theory in the limit of large number of open
 channels.  
\end{abstract}

\maketitle

\section{Introduction}

A quantum point contact (QPC) is one of the reference systems of
mesoscopic physics. The experimental discovery of conductance
quantizaton \cite{Expqpc} triggered further research which contributed
much to our modern understanding of nanoscience. QPC is a constriction
defined in a 2DEG by gates. The width of the constriction can be
changed by the voltage applied to these gates. In the adiabatic
regime, if the distance between the gates changes slowly compared to
the wavelength of an electron, the theoretical description is readily
obtained \cite{Glazman,Buttiker}. The two-dimensional motion of an
electron confined between the gates is equivalent to {\em
one-dimensional} scattering of an electron at a potential barrier. The
height of the barrier is different for different transport
channels. Semi-classically, the electron is fully transmitted if its 
energy exceeds the top of the barrier in a given channel, and is fully
reflected otherwise. Thus, semi-classical transmission eigenvalues of
a QPC are strongly degenerate: One has a finite number of transmission
eigenvalues equal to one and an infinite number of transmission
eigenvalues equal to zero. This picture would manifest experimentally
in the precise quantization of conductance as a function of gate
voltage. 

In real experiments, this degeneracy is lifted. A brief glance at any
of many available experimental studies shows that conductance does not
rise in ideal steps. The question whether the transmission eigenvalues
are degenerate is also important for a number of other reasons. For
instance, if a QPC is prepared in a superconducting material, discrete
subgap (Andreev) states develop \cite{Andreev}. These states describe
quasiparticles localized around QPC. The number of these states equals
the number of transport channels, and their energies are expressed via
transmission eigenvalues $T_n$, $E_n = \Delta \sqrt{1 - T_n \sin^2
\varphi/2}$, with $\Delta$ and $\varphi$ being the superconducting gap
and the phase difference across the QPC. If the transmission
eigenvalues are degenerate, the Andreev levels are also
degenerate. Thus, any small perturbation would lift this degeneracy
and produce a number of states with very close energies. Such a
perturbation would then drastically affect properties of the system.  

An obvious candidate for this degeneracy lifting is quantum tunneling
across the top of the barrier. Indeed, for a given energy there is a
range of gate voltages when {\em one} transport channel has a
transmission eigenvalue between zero and one -- a partially open
channel. All other transmission eigenvalues are also modified by the
quantum tunneling: They get an exponentially small correction. Thus,
quantum tunneling leads to the rounding of the conductance steps as a
function of gate voltage, but only provides exponentially small
splitting of Andreev states.  

In this paper, we study how the degeneracy of transmission eigenvalues
is lifted by the scattering on impurities, which are always present in
and around the QPC. Properties of a disordered QPC have been
investigated (see
Refs. \onlinecite{Imry,Nixon,Glazman2,Maslov,Beenakker}), mostly in
relation to the disorder smearing of conductance steps or evolution of 
conductance fluctuations in ballistic regime. In contrast to the
previous literature, we investigate the case when the conductance of
the QPC is only slightly modified by the impurities, or, in other
words, the impurity-related splitting of transmission eigenvalues is
much less than one. This regime is realized for low concentration of
impurities. In this situation we can disregard quantum
effects like resonant tunneling through impurity states or Kondo
effect. 

Our main result is that in this regime, reflection amplitudes are
Gaussian distributed with zero average and second-order correlation
function which does not depend on the channel index. This
provides us with a new class of random matrix theory. The results for
the distribution function of transmission eigenvalues are universal
--- they only depend on the number of transport channels and on the
average reflection eigenvalue. All other information can be
extracted from these two parameters. 

The paper is organized in the following way. In Section \ref{model}
we treat a disordered QPC in the adiabatic approximation. In
Section \ref{correl} we introduce the scattering matrix and show that
in the expansion up to the second order in disorder potential closed
channels do not contribute to the properties of transmission
eigenvalues of open channels. Section \ref{scat} finalizes the
quantum-mechanical calculation of reflection coefficient and
conductance of a disordered QPC. We then turn to the classical
(Boltzmann equation) consideration, which facilitates the
consideration of the diffusive regime (Section \ref{Boltzmann}). 

In Section \ref{noise} we discuss noise properties of disordered
QPC. Finally, Section \ref{distrib} is devoted to the distribution
function of transmission eigenvalues. For one open transport channel, we
calculate this distribution function analytically by performing the
disorder averaging directly. In the limit of large number of open
channels, we obtain the distribution function by means of the
circuit theory \cite{Nazarov}, which presents a disordered QPC as a
pure QPC and a diffusive resistor connected in series. For
intermediate numbers of open channels, we perform a numerical
simulation based on random matrix theory.  

\section{Model of QPC with impurities} \label{model}

We describe the QPC as a constriction between two infinitely high
walls \cite{Glazman} separated by the distance $d(x)$ (Figure 1). A
more physical model would take into account that the transverse
profile is not sharp \cite{Buttiker}. Since in this paper we employ
semi-classical approximation (do not discuss the rounding of
conduction steps), the results do not depend on the details of the
potential profile. For this reason, we use the simpler model. The
Schr\"{o}dinger equation,
\begin{equation}\label{s1}
\left[ -\frac{\hbar^2}{2m} \nabla^2 +V(x,y)
\right]\psi(x,y)=E\psi(x,y), 
\end{equation}
is supplemented by the boundary conditions,
\[
  \psi(x, y = {\pm}d(x)/2)=0.
\]
Here
\[V(x,y)=\sum_i \, v(x-x_i,y-y_i),
\]
with $v$ being the single impurity potential, and the sum is taken
over impurity positions.  

If the width of the constriction $d(x)$ changes smoothly, we can
employ the adiabatic approximation and separate the transverse motion, 
\[ \psi(x,y) = \sum_n \phi_n(x) \varphi_n^{(x)}(y). \]
The transverse wave functions $\varphi_n^{(x)}(y)$ that satisfy the
boundary conditions are
\[ \varphi_n^{(x)}(y)=\sqrt{\frac{2}{d(x)}} \sin \left[\frac{n\pi
}{d(x)}\, 
\left(y+\frac{d(x)}{2}\right)
\right]. \]
Substituting this into Eq. (\ref{s1}) and disregarding the terms
containing the derivatives of $d(x)$, we obtain a {\em
one-dimensional} equation for the longitudinal wave function,
\begin{equation}\label{s2}
\left[ -\frac{\hbar^2}{2m}\frac{d^2}{dx^2}+\epsilon_n(x)-E\right]
\phi_n(x)=-\sum_m V_{nm}(x)\phi_m(x),
\end{equation}
with the channel-dependent effective potential barrier
\[ \epsilon_n(x)=\frac{\hbar^2\pi^2n^2}{2 m d^2(x)},\]
and the matrix element of the disorder potential,
\[ V_{nm}(x)=\int_{-d(x)/2}^{d(x)/2} dy \, \varphi_n^{(x)}(y) V(x,y)
\varphi_m^{(x)}(y).\]
Eq. (\ref{s2}) is the generalization of the equations previously
written in Refs. \onlinecite{Glazman,Buttiker} to the case of
disordered QPC.

In the semi-classical (WKB) approximation, in the absence of disorder,
for each transport channel, electrons with the energies above
(below) the top of the barrier are perfectly transmitted
(reflected). This approximation breaks down if the energy of an 
electron coincides with the top of the barrier. In this paper, we do
not consider this case. The wave function of an ideally
transmitted electron is 
\begin{equation}\label{w1}
 \phi_n^{(0)}(x)=\sqrt{ \frac{p_n(\infty)}{p_n(x)} } \exp
\left[ \frac{i}{\hbar} \int^x p_n(z) dz \right],
\end{equation}
with the channel-dependent momentum
\[p_n(x)=[2m(E-\epsilon_n(x))]^{1/2}.\]

\section{Correlators of the scattering matrix elements} \label{correl}

We proceed by introducing the scattering matrix, 
\[
\hat S=\left(
\begin{array}{cc}
\hat r & \hat t \\ \hat t^T & \hat r'
\end{array}
\right),
\]
which is unitary, ${\hat S}^\dag {\hat S}=1$, due to the current
conservation requirement. At zero temperature, conductance of the
system is expressed via Landauer formula, 
\[
G = G_Q \mbox{Tr}\, {\hat t}^\dag {\hat t} = G_Q
\sum_n T_n,
\]
where $T_n$ (of interest in this paper) are the eigenvalues
of the matrix $\hat t^{\dagger} \hat t$, and $G_Q = e^2/\pi\hbar$ is
the conductance quantum. Without impurities, the matrix $\hat
t^{\dagger} \hat t$ is diagonal, with the elements describing the
transmission of an electron in the same open transport channel equal
one and all others equal zero. In this case, the conductance is $G_0 =
G_QN$, with $N$ being the number of open transport channels. 

To treat the effect of disorder, it is more convenient to investigate
the matrix $\hat r^{\dagger} \hat r$, with the eigenvalues ({\em
reflection eigenvalues}) $R_n = 1 - T_n$. In the following, we
calculate the correction to the transmission eigenvalue $R_n$ due to
disorder. For this purpose, we consider the perturbation expansion of
the reflection matrix $\hat r$ up to the second order in the disorder
potential, 
\[\hat r=\hat r^{(0)}+\hat r^{(1)}+\hat r^{(2)}. \]
Let us now separate open and closed channels, 
\[
\hat r=\left(
\begin{array}{cc}
\hat r_{oo} & \hat r_{oc} \\ \hat r_{co} & \hat r_{cc}
\end{array}
\right),
\]
where the submatrices $r_{\alpha\beta}$, $\alpha, \beta = o,c$
describe reflection from/to channels of different type ($o$ and $c$
stand for open and closed). The reflection eigenvalues are found from
the secular equation, $\det \left( {\hat r}^\dag{\hat r}- R {\hat 1}
\right) =0$, or, equivalently,  
\begin{equation} \label{secular2} 
\det \left( \begin{array}{cc} \hat r^\dag_{oo} \hat r_{oo} + \hat
r^\dag_{co} \hat r_{co} - R {\hat 1}
& \hat r^\dag_{oo} \hat r_{oc} + \hat r^\dag_{co} \hat r_{cc}\\
\hat r^\dag_{oc} \hat r_{oo} + \hat r^\dag_{cc} \hat r_{co} &
\hat r^\dag_{oc} \hat r_{oc} + \hat r^\dag_{cc} \hat r_{cc} - R {\hat
1} \end{array} \right) = 0.
\end{equation}
We now expand this equation in powers of the disorder potential. For
{\em open channels}, the reflection eigenvalues $R$ are expected to be
of the second order in disorder. We use now the identity 
\[
\det \left( \begin{array}{cc} A & B \\
C & D \end{array} \right) = \det (A-BD^{-1}C) \det D,
\]
and expand the first determinant taking into account that
$\hat r^{(0)}_{oc} = \hat r^{(0)}_{co} = \hat r^{(0)}_{cc} = 0$. The
terms of zeroth order in the disorder potential cancel, since $\hat
r^{(0)\dag}_{cc} \hat r^{(0)}_{cc} = 1$. Terms of the first order do
not appear, and in the second order one has $\det ( \hat
r^{(1)\dag}_{oo}r^{(1)}_{oo} - R{\hat 1}) = 0$. Thus, closed
channels have no effect on transmission of open channels. In the rest 
of the paper, we drop the subscript $oo$ and operate only with the
quantities related to open channels. 

Elements of the matrix $\hat r$ are random quantities. In the next
Section, we characterize their statistical properties. We show that
they are Gaussian distributed with zero average. Thus, it is enough to
specify the correlation functions
\[ \langle {\hat r}_{ij}^{*(1)} {\hat r}_{kh}^{(1)} \rangle,\]
where the average is performed with respect to disorder. We show that
the only non-zero correlator is $\left\langle\vert r^{(1)}_{ij}
\vert^2\right\rangle$, which is the probability for an electron coming
in the open channel $i$ to be scattered to the open channel $j$. As a
matter of fact, the result does not depend on the indices $i$ and
$j$. We then relate this reflection probability with the correction to
the conductance.  

Thus, the set of eigenvalues $R_n$ describing {\em open channels} is
obtained by diagonalizing the {\em finite-size} random matrix $\hat
r^{\dag} \hat r$. This matrix is Gaussian, and depends only on one
parameter --- average reflection coefficient. This novel random matrix
theory in later Sections provides a novel distribution of
transmission eigenvalues. 

\begin{figure} \label{fig1}
\includegraphics[scale=0.3]{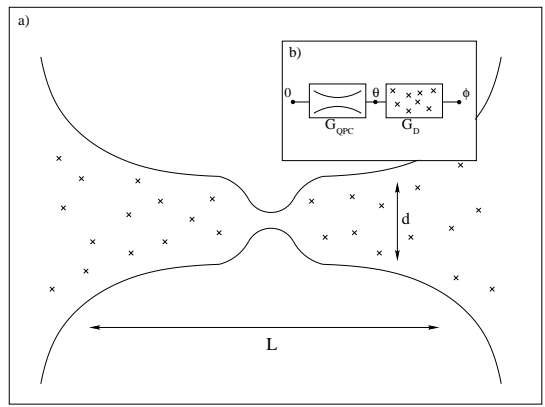}
\caption{(a) Layout of a disordered QPC. (b) An equivalent circuit
representing a disordered QPC as a clean QPC and a diffusive resistor
in series.}
\end{figure} 

\section{Scattering matrix approach} \label{scat}

To calculate the  correction to the reflection amplitudes we consider
the Green's functions of Eq. (\ref{s2}),
\[
\left[ -\frac{\hbar^2}{2m}\frac{d^2}{dx^2}+\epsilon_n(x)-E\right]
{\mathcal G}_n(x,x')=-\delta(x-x').\]
Solving this equation, we find the Green's functions for open
channels, 
\begin{equation}\label{g1}
 {\mathcal G}_n(x,x') =\frac{i\hbar}{m\sqrt{p_n(x)p_n(x')}}
\exp\left[ \frac{i}{\hbar}
\left| \int_x^{x'} dz  \, p_n(z) \right| \right].
\end{equation}
The formal solution of Eq. (\ref{s2}) takes the form
\begin{equation}\label{extsol}
\phi_n(x)=\phi_n^{(0)}(x)+ \int dx' {\mathcal G}_n(x,x') \sum_m
V_{nm}(x')\phi_m(x'),
\end{equation}
with $\phi_n^{(0)}$ being the solutions in the absence of disorder
(Eq. (\ref{s2}) with the zero right-hand side.) In the first order in
$V$, we obtain 
\begin{equation}\label{dyson}
\phi_n(x)=  \phi_n^{(0)}(x)+\sum_m\int dx' {\mathcal G}_n(x,x')
V_{nm}(x')\phi^{(0)}_m(x').
\end{equation}
Substituting Eqs. (\ref{w1}) and (\ref{g1}) into Eq. (\ref{dyson}), we
find
\begin{eqnarray} \label{reflcoeff}
r_{nm} & = & \int_{- \infty}^\infty dx'\frac{i
\hbar}{m\sqrt{p_n(x')p_m(x')}} \nonumber \\
& {\times} &  \exp\left[ \frac{i}{\hbar}\int^{x'}dz \left( p_n(z)+p_m(z)
\right) \right] V_{nm}(x').
\end{eqnarray}
For the Gaussian distribution of disorder, the reflection amplitudes
$r_{nm}$ are also Gaussian distributed. This distribution is fully
characterized by the pair correlation function. If the impurities are
not located in the constriction, the momenta $p_n (x)$ and $p_m(x)$ in
Eq. (\ref{reflcoeff}) can be replaced by their values taken at $x \to
\infty$, which is $p_F$ independently of the channel index. The
impurity averaging is straightforward. As anticipated, all the
averages of the type $\langle {\hat r}_{ij}^{*} 
{\hat r}_{kh} \rangle$ turn to zero due to the oscillating behavior,
except for the term $\langle \vert r_{nm} \vert \rangle^2$,  
\begin{equation} \label{correl1}
\langle \vert r_{nm} \vert^2 \rangle = \frac{n_i}{\hbar^2
v_F^2}\frac{L}{d} \left\vert {\tilde v} (\pi) \right\vert^2
\left( 1+\frac{\delta_{nm}}{2} \right),
\end{equation}
where ${\tilde v} (\pi)$ is the Fourier transform of the single
impurity potential with the momentum transfer $2k_F$, and $n_i$ is the
concentration of impurities per unit area.

The correction to the conductance reads
\[
\langle \delta G \rangle =-G_Q\sum_{n,m=0}^N\langle |r_{nm}|^2 \rangle,
\]
and in the case of large number of open channels $N$ can be written as
\begin{equation}\label{g2}
\langle \delta G \rangle \simeq -\frac{e^2 n_i}{\pi \hbar^3
v_F^2}\frac{L}{d}|{\tilde v}(\pi)|^2 N^2.
\end{equation}

For further reference, we identify the average reflection eigenvalue
$\langle R \rangle$ by means of Landauer formula, $\langle \delta G
\rangle = - G_Q N \langle R \rangle$,
\begin{displaymath}
\langle R \rangle = \frac{n_i}{\hbar^2
v_F^2}\frac{L}{d}|{\tilde v}(\pi)|^2 N.
\end{displaymath}
The correlation function of the reflection amplitudes can then be
expressed via only one parameter $\langle R \rangle$,
\begin{equation} \label{correl10}
\langle \vert r_{nm} \vert^2 \rangle = \frac{1}{N} \left\langle R
\right\rangle \left( 1+\frac{\delta_{nm}}{2} \right).
\end{equation}

\section{Boltzmann equation} \label{Boltzmann}

In this Section we analyse transport properties of a disordered
QPC in the framework of the Boltzmann equation. This approach allows
for an extension of the analysis to the diffusive case, when the mean
free path becomes much smaller then the length of the system. In
this case the second order perturbation expansion breaks down, and the
treatment of previous Sections can not be applied any more. 

We model the system as a two-dimensional disordered wire between ideal
reservoirs. To take into account the constriction, we recall that
without disorder, in the quantum treatment only the channels with low
index are ideally transmitting (open). In the language of classical
physics, these channels correspond to electronic modes propagating
with small incident angle $\alpha$ (Figure 2). To implement this
feature in our model, we assume that all electrons propagating with
angles smaller (larger) then $\alpha_0 \ll 1$ are perfectly
transmitted through the cross-section $x=0$ (to model the position of
QPC). Electrons with higher angles are reflected from this
cross-section. The Boltzmann equation reads
\begin{equation} \label{bolt1} {\bm v} {\cdot} {\bm \nabla_r}f({\bm
r},{\bm p} )=I[f], 
\end{equation}
where $f({\bm r},{\bm p} )$ is the distribution function of electrons,
and $I[f]$ is the collision integral \cite{abrikosov},
\[ I[f] = \frac{2\pi n_i}{\hbar} \int \frac{d^2{\bm p'}}{(2 \pi \hbar)^2}
[f({\bm r,\bm p}')-f({\bm r,\bm p})] \]
\[ {\times}|{\tilde v}({\bm p-\bm p'})|^2 \delta \left( \epsilon({\bm
p}')-\epsilon({\bm p}) \right). 
\] 
Since only electrons with energies betwen $E_F$ and $E_F + eV$, with
$V$ being the applied voltage, contribute to the net current, the
absolute value of the momentum $p$ is fixed to lie at the Fermi
surface. We are only interested in the angular dependence of the
distribution function, $f(x,\alpha)$. The Boltzmann equation
(\ref{bolt1}) is supplemented by the boundary conditions, 
\[ \left\{
\begin{array}{lr}
f(x,\alpha)=1, & \cos \alpha > 0, x < 0 \\ f(x,\alpha)=0, &
\cos \alpha < 0, x > L ,
\end{array}
\right. 
\]

which state that electrons coming from the reservoirs are in thermal
equilibrium. To take into account reflection of electrons at the QPC,
we introduce the further boundary condition,
\[
f(0,\alpha) = f(0,\pi-\alpha), \ \ \ \ \ \vert \cos \alpha \vert <
\cos\alpha_0. 
\]
\begin{figure} \label{fig2}
\includegraphics[scale=0.3]{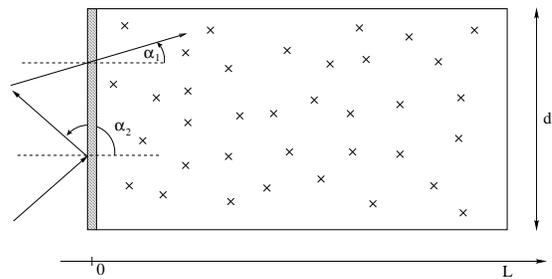}
\caption{Disordered QPC modeled classically. Electrons with the angle
$\alpha$ lower than the critical angle $\alpha_0$ are transmitted
through the cross-section $x=0$, others are reflected from this
cross-section.}
\end{figure}

The distribution function does not depend on the transverse coordinate
$y$, and we thus rewrite Eq. (\ref{bolt1}) as
\begin{eqnarray}\label{bolt2}
& & v_F \cos \alpha \ \partial_x f(x,\alpha) \\
& = & \frac{n_i m}{2 \pi \hbar^2 }
\int d\alpha '\cos \alpha' \left\vert {\tilde v}
(\alpha-\alpha') \right\vert^2 \left[ f(x,\alpha) - f(x,\alpha ')
\right]. \nonumber 
\end{eqnarray}
We solve now Eq. (\ref{bolt2}) in the two limiting cases of low and
high impurity concentration. The first case corresponds to the
consideration of previous Sections based on the scattering
approach. The second one, instead, allows us to investigate the
diffusive regime. 

\subsection{Non-diffusive regime}

For low impurity concentration, we replace the function $f(x,
\alpha')$ in the rhs of Eq. (\ref{bolt2}) with the distribution
function of the pure system. The remaining differential equation is
readily solved, yielding for the distribution function at $x=0$ and
with the direction of propagation $\vert \cos \alpha \vert > \cos
\alpha_0$,   
\[
f(0,\alpha) = \left\{
\begin{array}{ll}
1, & \cos \alpha >0 \\
\displaystyle -\frac{\alpha_0 L n_i m |{\tilde
v} (\alpha)|^2}{v_F \pi \hbar^3 \cos \alpha},
& \cos \alpha<0
\end{array} \right. .
\]
Now we calculate the current density, ${\bm j} = (e\nu(E_F) v_F/2\pi)
\int d\alpha \cos \alpha f(0, \alpha)$, with $\nu (E_F)$ being the
density of states at the Fermi energy. Only the angles $\vert \cos
\alpha \vert > \cos \alpha_0$ contribute, and we obtain the
conductance, 
\begin{equation}
G=e^2 v_F \nu(E_F) \left[2 \alpha_0 -2\alpha_0^2
\frac{n_im}{2\pi \hbar^3v_F}L|{\tilde v}(\pi)|^2\right]d,
\end{equation}
where we have taken into account that $\alpha_0 \ll 1$. The second
term represents the correction to the conductance due to impurity
scattering. Relating the number of open channels to the critical angle
$\alpha_0$, $\alpha_0 = 2N \pi / k_F d$, we find that this correction
is identical to Eq. (\ref{g2}), which we obtained from the scattering
matrix approach in the limiting case $N \gg 1$. 

\subsection{Diffusive regime}

For high impurity concentration, we introduce the transport relaxation
time $\tau$, defined as \cite{abrikosov}
\[\tau^{-1}=\frac{n_i m}{2\pi \hbar^3}\int_0^{2\pi}d\alpha
|\tilde v (\alpha)|^2 (1 - \cos \alpha).
\]
In the linear regime, the collision integral takes the form $I[f] = - 
(f- {\bar f})/\tau$, where $\bar f$ is the distribution function 
averaged over the angles, 
\[{\bar f}=\frac{1}{2\pi} \int_0^{2\pi} d \alpha f(\alpha).
\]
Solution of the Boltzmann equation in this case is given by
\[f(x,\alpha)=\left(\frac{\pi l}{2L\alpha_0}+1\right)^{-1}
\left(1-\frac{x}{L} 
+\frac{l}{L}\cos \alpha \right),
\]
with $l= v_F \tau $ being the mean free path, $l \ll L$. Calculating
the current, we find that the conductance is 
\begin{equation} \label{Boltzcond}
G= \frac{e^2 v_F \nu(E_F) d}{2\pi} \left(
\frac{1}{2\alpha_0}+\frac{L}{\pi l} 
\right)^{-1}. 
\end{equation}
Identifying the conductance of the diffusive region,
\[
G_{D}=\frac{e^2 v_F \nu(E_F) l d}{2L},
\]
and of the clean QPC,
\[
G_{QPC}=\frac{e^2 v_F \nu(E_F) d}{2\pi} 2 \alpha_0 = G_QN, 
\]
we see that Eq. (\ref{Boltzcond}) represents a series addition of the
conductances of the clean QPC and a diffusive resistor.

\section{Noise} \label{noise}

In a two-terminal system, zero-temperature shot noise \cite{BB} can be
expressed in terms of reflection eigenvalues in the following way, 
\[
{\mathcal S} = \frac{e^3|V|}{\pi \hbar} \left\langle \sum_n R_n(1-R_n)
\right\rangle.
\]
In a clean QPC, all transmission eigenvalues in the semi-classical
regime are either zero or one, and the current is noiseless. Quantum
tunneling away from the conductance steps only brings exponentially
small contribution. Thus, we expect a drastic effect of disorder on
the shot noise of a QPC. Indeed, up to the second order in the
disorder potential, one writes
\begin{equation} \label{noiseqpc}
{\mathcal S} = - \frac{2e^3 |V|}{\pi \hbar}\sum_n \delta R_n = - 2e
\vert V \vert \delta G = 2eG_Q N \vert V \vert \langle R \rangle.  
\end{equation}
Eq. (\ref{noiseqpc}) corresponds to the notion of Poissonian stream of 
reflected particles, which is, indeed, expected in the case of good
transmission.  

\section{Distribution function} \label{distrib}

Now, we turn to the calculation of the distribution function of
transmission eigenvalues, 
\[\rho(T)=\left\langle \sum_n \delta(T-T_n) \right\rangle,
\]
where the sum is taken over all open channels, and disorder averaging
is performed. The distribution function is normalized so that its
integral is the number of open channels $N$. For a clean QPC, $\rho
(T) = N \delta (1-T)$. 

We first consider the case of one open channel and perform the
disorder averaging directly. Then, we obtain analytical results in the
limiting case of very large number of channels based on the circuit
theory: A disordered QPC is presented as a clean QPC connected in
series with a diffusive resistor. For intermediate values of $N$, we
were not able to obtain analytical results. Instead, we perform a
simulation based on the notion of random reflection matrices. 

\subsection{One open channel}

For one open channel, we perform a ``brute force'' disorder
averaging. The reflection probability is $R = \vert r \vert^2$, where
we have suppressed the channel indices. In its turn, the reflection
amplitude $r$ is related to the potential $V$ by
Eq. (\ref{reflcoeff}), with $m=n=1$. It is a complex
quantity, and we first calculate the distribution function of the
transmission amplitude, defined as 
\[ 
{\mathcal P}(r) = \left\langle \delta \left( \mbox{Re} ( r- r[V])
\right) \delta \left(\mbox{Im} (r- r [V]) 
\right) \right\rangle.   
\]
Representing delta-functions as integrals, we write explicitly
\begin{eqnarray} 
& & {\mathcal P}(r) = \left\langle
\int_{-\infty}^{\infty}\frac{d\omega_1 d\omega_2}{(2\pi)^2} \right.
\nonumber \\
& \times & \left. \exp\left\{ i\omega_1\mbox{Re}\left[ r[V]-r\right] + 
i\omega_2\mbox{Im}\left[ r[V]-r\right] \right\} \right\rangle .
\end{eqnarray}
Performing the averaging with the Gaussian distribution
$\tilde {\mathcal P} [V]$, and introducing the short-hand notation
\begin{displaymath} 
r[V]=\int d{\bm r} F({\bm r})
V({\bm r}),
\end{displaymath}
we obtain 
\begin{widetext}
\begin{eqnarray*} 
& & {\mathcal P}(r) = \int_{-\infty}^{\infty} \frac{d\omega_1
d\omega_2}{(2\pi)^2} \exp\left\{-\frac{1}{2}\int d{\bm r_1}d {\bm r_2} 
\left [ \omega_1\mbox{Re} F({\bm r_1}) + \omega_2 \mbox{Im} 
F({\bm r_1}) \right] \left\langle V({\bm r_1}) V ({\bm r_2})
\right\rangle \left[
\omega_1\mbox{Re} F({\bm r_2}) + \omega_2\mbox{Im} F({\bm r_2})
\right] \right. \nonumber \\ 
& & \left. -i\omega_1 \mbox{Re}\,
r-i\omega_2
\mbox{Im}\, r  \right\}.
\end{eqnarray*}
\end{widetext}
Now we use Eq. (\ref{reflcoeff}) and disregard the terms containing
rapidly oscillating functions. Calculating the integrals over
$d\omega_1$ and $d\omega_2$, we finally obtain  
\begin{equation} \label{distr1}
\mathcal P(r)=\frac{1}{\pi \langle
|r|^2\rangle}\exp\left(-\frac{|r|^2}{\langle|r|^2\rangle }\right). 
\end{equation}
Since $dR = 2 \vert r \vert d \vert r \vert = \pi^{-1} d\ {\mbox Re} 
r\ d\ \mbox{Im} r$, we can rewrite Eq. (\ref{distr1}) in terms of the
reflection eigenvalue,
\begin{equation} \label{distr2}
\rho(R) = \frac{1}{\langle R \rangle}\exp
\left(-\frac{R}{\langle R \rangle}\right); \ \ \ \rho(T) = \rho
(1-R). 
\end{equation}
Thus, for one channel the reflection eigenvalue is Poisson
distributed. This result is actually not surprising. Indeed, both real
and imaginary parts of $r$ are linearly related to the potential
$V$. This means they are both Gaussian with zero
average. Moreover, from Eq. (\ref{reflcoeff}) it follows that they
have the same dispersion, and we arrive to the Gaussian distribution
for $r$ (\ref{distr1}) and Poisson distribution for $R$
(\ref{distr2}). These distributions are {\em universal}: All the
information about the type and amplitude of disorder is encoded in
only one number, which is the average reflection eigenvalue $\langle R
\rangle$. 

\subsection{Circuit Theory}

If the number $N$ of open channels is large ($G\gg G_Q $), we can
calculate the distribution function $\rho(T)$ analytically by means of
the circuit theory developed by one of the authors \cite{Nazarov}. To
this purpose, we represent a disordered QPC as a clean QPC connected
in series to a diffusive conductor (Figure 1). Such a point of view,
to our knowledge, was first adopted in Ref. \onlinecite{Maslov} for
investigation of conductance fluctuations. The input parameters
are the conductances of both circuit elements (connectors), $G_{QPC}$
and $G_D$. Each connector is subject to a phase difference $\phi$,
which generates the pseudo-current $I(\phi)$. The relation $I(\phi)$
is determined by the distribution function of transmission
eigenvalues, 
\[ 
\rho(T)\,=\, \frac{1}{2\pi G_Q}\frac{1}{T\sqrt{1-T}} \mbox{Re}
\left[ I \left( \pi + 2 i \mbox{arccosh} \frac{1}{ \sqrt{T} } \right)
\right]. 
\]
In our case, the current-phase relations are 
\begin{itemize}
\item Diffusive conductor: $I(\phi)= G_D \phi$
\item Quantum point contact $I(\phi)=2 G_{QPC} \tan(\phi /2)$
\end{itemize}
The circuit theory shows how these two elements can be combined to get
the distribution function of the entire circuit. To this purpose, we
introduce the phases: $\phi$ in the left reservoir; zero in the right
reservoir, and $\theta$ (to be calculated) in the point (node)
separating the QPC and diffusive conductor. Thus, the QPC is subject
to the phase difference $\phi - \theta$, and the diffusive conductor
to the phase difference $\theta$. The pseudo-current must be
conserved, from which we get the following equation for $\theta$, 
\begin{equation} \label{circuitcur}
G_D \theta = 2 G_{QPC} \tan ((\phi - \theta)/2). 
\end{equation}
After solving this equation, we find the pseudo-current $I (\phi) =
G_D \theta (\phi)$, and eventually the distribution function. 

Eq. (\ref{circuitcur}) can not be solved analytically for an arbitrary
relation between $G_D$ and $G_{QPC}$, and we restrict ourselves to the
case of low impurity concentration, $G_D \gg G_{QPC}$. We obtain 
\[ I(\phi)\,=\, G_D \left[ \frac{\phi-\pi}{2}+
\sqrt{\frac{(\phi-\pi)^2}{4}+\frac{4 G_{QPC}}{G_{D}}}\right],
\]
and the distribution function follows, 
\begin{equation} \label{circuitdistrib}
\rho(T)=
\frac{G_D}{2 \pi G_{Q}}\frac{1}{T\sqrt{1-T}} \sqrt{T-T_c},
\end{equation}
for $T > T_c$,  $T_c= 1 - 4 G_{QPC}/G_D$, and zero for $T < T_c$. For
further comparison with the simulation results, we calculate the
average reflection coefficient $\langle R \rangle = G_{QPC}/G_D \ll 1$
and rewrite the distribution function (\ref{circuitdistrib}) as
\begin{equation} \label{circuitdistrib1}
\rho(R)= \frac{N}{2\pi \langle R \rangle} \frac{1}{1-R}
\sqrt{\frac{4\langle R \rangle}{R} - 1}. 
\end{equation}
We see that in the limiting case $N \gg 1$ only channels with the
transmission close to perfect exist: Reflection eigenvalues can only
be lower than $4 \langle R \rangle$, which is a small number. This is
in contrast to the one-channel case, where all values of the
reflection coefficient are permitted. To demonstrate the crossover
between these two limiting cases, we perform a numerical
simulation for the intermediate number of open transport channels.  

\begin{figure}[ht] \label{fig3}
\includegraphics[angle=0,width=8 cm]{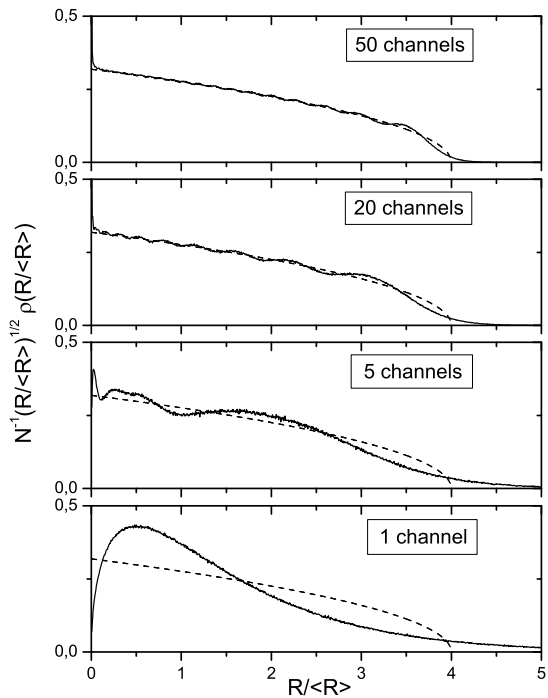}
\caption{The distribution function of reflection eigenvalues $\rho(R)$ for
1, 5, 20 and 50 open channels. It is calculated using an ensamble of $10^6$
random matrices. The dashed line is the  expression
(\ref{circuitdistrib1}) valid in the limit $N \rightarrow \infty$.}
\label{fig:QPC}
\end{figure}

\subsection{Intermediate case}

For any number of open channels, the matrix elements of the
reflection matrix, $r_{nm}$ are random quantities with Gaussian
distribution. They are zero on average, uncorrelated, and 
characterized by the dispersion $\langle \vert r_{nm} \vert^2
\rangle$. Eq. (\ref{correl1}) shows that this dispersion is with the
factor of $3/2$ greater for diagonal elements that for non-diagonal
elements, and otherwise does not depend on $n$ and $m$. Then, the
problem has only two parameters --- the number of open transport
channels $N$, and the dispersion of matrix elements of the reflection
matrix. The latter parameter is related to the average reflection
eigenvalue. 
 
Thus, we model the reflection matrix as a Gaussian random matrix with 
independent elements $r_{nm}$. Both real and imaginary part of
$r_{nm}$ are taken to be Gaussian distributed with the same
dispersion (independent on $n$ and $m$). The simulation involves
$10^6$ matrices. Results for $1$, $5$, $20$, and $50$ channels are
plotted in Figure 3. For better comparison with
Eq. (\ref{circuitdistrib1}), we multiply the distribution function
with $R^{1/2}$, so that it tends to a constant value at $R \to 0$. For
one channel, the simulation result perfectly reproduces the Poisson
distribution (\ref{distr2}), whereas for $50$ channels it is in a good
agreement with Eq. (\ref{circuitdistrib1}). For $5$ and $20$ channels,
the simulations reveal an intermediate picture, with a tail at large
$R$. Oscillations in the distribution function, pronounced for $5$,
$20$, and $50$ channels, are related to the Wigner-Dyson correlations
of eigenvalues of Gaussian random matrices. 

\section{Conclusions}

The main result of our paper is the distribution function of
reflection eigenvalues of a disordered QPC. We have shown that for one
open channel, one obtains Poisson distribution, and in the limit of
infinitely many channels only very small reflection eigenvalues are
allowed. We also performed numerical simulations for the intermediate
regime. In all the cases, the results are universal in the sense that
they only depend on the number of open transport channels and on the
average reflection coefficient $\langle R \rangle$. To calculate
$\langle R \rangle$, we developed a quantum-mechanical theory, and a
classical calculation based on the Boltzmann equation. For low
impurity concentration, classical and quantum-mechanical results are
identical; additionally, classically one can treat the diffusive
regime, to find that the conductance of a disordered QPC is given as a
series resistance addition. We stress that although the results for
$\langle R \rangle$ are model dependent, as carefully investigated
previously by Glazman and Jonson \cite{Glazman2}, the form of the
distribution function only uses $\langle R \rangle$ as a parameter,
but does not depend on specific model any more. It can be regarded
as a result of the novel random matrix theory with Gaussian reflection 
matrices. 

This work was supported by the Netherlands Foundation for Fundamental
Research on Matter (FOM).

\end{document}